# Compressed Crystalline Bismuth and Superconductivity−An *ab initio* computational Simulation

David Hinojosa-Romero[1], Isaías Rodríguez[2], Zaahel Mata-Pinzón[1], Alexander Valladares[2], Renela Valladares[2], and Ariel A. Valladares[1] *

[1]Instituto de Investigaciones en Materiales, Universidad Nacional Autónoma de México, Apartado Postal 70-360, Ciudad Universitaria, México, D.F. 04510, México.

[2]Facultad de Ciencias, Universidad Nacional Autónoma de México, Apartado Postal 70-542, Ciudad Universitaria, México, D.F. 04510, México

*valladar@unam.mx

## ABSTRACT

Bismuth displays puzzling superconducting properties. In its crystalline equilibrium phase, it does not seem to superconduct at accessible low temperatures. However, in the amorphous phase it displays superconductivity at ~ 6 K. Under pressure bismuth has been found to superconduct at $T_c$s that go from 3.9 K to 8.5 K depending on the phase obtained. So the question is: what electronic or vibrational changes occur that explains this radical transformation in the conducting behavior of this material? In a recent publication we argue that changes in the density of electronic and vibrational states may account for the behavior observed in the amorphous phase with respect to the crystal. We have now undertaken an *ab initio* computational study of the effects of pressure alone maintaining the original crystalline structure and compressing our supercell computationally. From the results obtained we infer that if the crystal structure remains the same (except for the contraction), no superconductivity will appear.

## INTRODUCTION

Bismuth (Bi) is an interesting material for besides having a very high thermal conductivity, this semimetal displays puzzling superconducting properties. In its crystalline equilibrium phase it does not seem to display superconductivity at low temperatures. However, in the amorphous phase it displays superconductivity at ~ 6 K [1]. Under pressure bismuth has been found to superconduct at 2.55 GPa (Bi-II monoclinic crystalline phase), 2.7 GPa (Bi-III tetragonal phase) and 7.7 GPa (Bi-V body centered cubic phase), having superconducting transition temperatures of $T_c$ = 3.9 K, 7.2 K and 8.5 K, respectively [2]. Therefore, it is desirable to investigate what changes in the electronic or vibrational properties occur that may explain this radical transformation in the conducting behavior of this material. In a recent publication [3] we argue that changes in the density of electronic and vibrational states may account for the behavior in the amorphous phase. That is why we have carried out an *ab initio* computational study of the effects of structural changes when crystalline Bi is subjected to pressure. In order to see the effect of pressure alone we maintain the original crystalline structure and computationally compress our sample, a 64–atom supercell with periodic boundary conditions, between 0 and 15

GPa. We then calculate the electronic density of states, eDoS, and the vibrational density of states, vDoS. We shall present a detailed discussion and infer how relevant the observed changes are to account for superconductivity.

The field of superconductivity has experienced an important growth and we speak of conventional and unconventional superconductors and also the ill-defined undetermined category [4]. Superconductivity in amorphous Bi is considered conventional and as such the Bardeen Cooper and Schrieffer (BCS) theory should be applicable. Since crystalline Bi has not been found to superconduct no categorization can be applied but it seems safe to assume that it would be conventional. Also, we may assume that the under-pressure superconducting phases found experimentally may be conventional. For these materials the BCS formula for the transition temperature $T_c$ should also apply:

$$T_c = 1.13\theta_D \exp[-1/N(E_F)V_0]$$

If we graph the dependence of the transition temperature on $\theta_D$, the Debye temperature, and on $N(E_F)$, the density of states at the Fermi level, we find the surface depicted in Figure 1. $V_0$ is the e-e phonon-mediated attraction. From this figure one can see that the increase in $T_c$ is more pronounced with $N(E_F)$ (exponential) than with $\theta_D$ (linear). This implies that if a given material undergoes a process that increases $N(E_F)$, the $T_c$ will increase more prominently than when the Debye temperature increases. This is consistent with recent findings of compressed $H_2S$ where large increases of $N(E_F)$ are found, increments that may lead to very high superconducting transition temperatures [5]. Also, the superconductivity in amorphous Bi seems to be due to a fourfold increase in the density of states, as described in reference 3. That is why we want to see how eDoS and vDoS change when pressure is applied to an otherwise invariant crystalline structure.

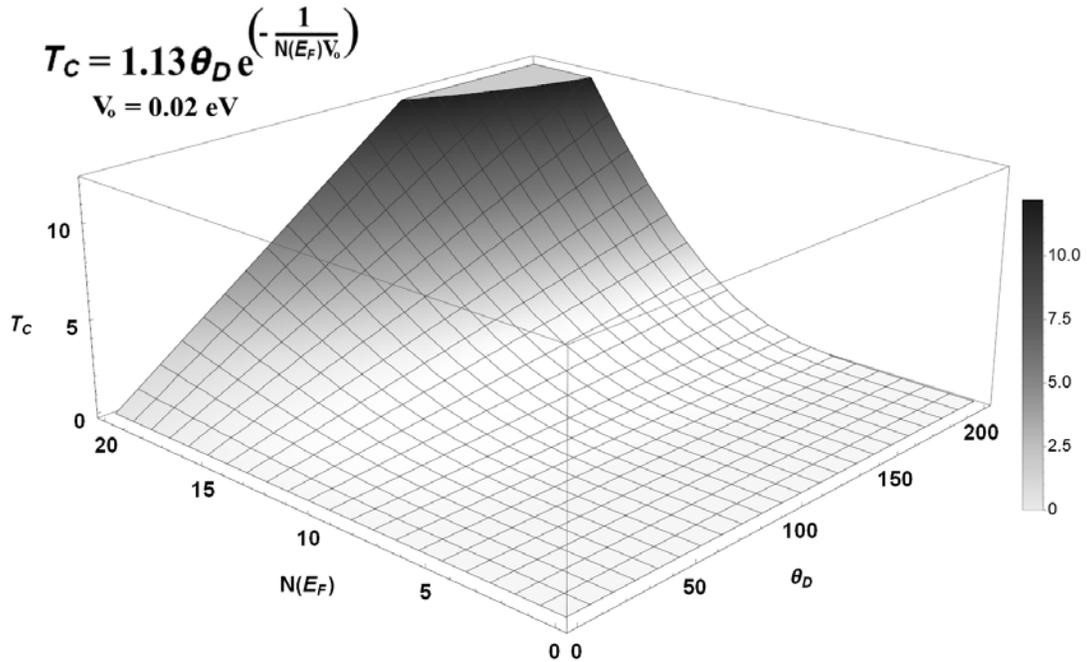

**Figure 1.** The BCS theory predicts a transition temperature that depends on the vibrational properties $\theta_D$ and the electronic properties $N(E_F)$. $V_0$ was chosen as 0.02 for our purposes.

What we mean by the crystalline structures are the structures that under compression maintain the symmetry properties constant like the crystalline angles of the original equilibrium cell. Clearly, the lattice parameters and the interatomic distances will change accordingly.

**THEORY**

In order to calculate the corresponding pressure for a given compression (percentage of change of volume) of the supercell we do the following. We begin with the structure reported by Wyckoff [6], construct a periodic supercell with 64 atoms, calculate its binding energy (BE) to a high precision ($10^{-6}$ in energy, 6 unique symmetry points in k-space, and 5 decimal places), find the parameters that make this structure a minimum of energy (stable structure) and with these parameters calculate 18 compressed and expanded periodic supercells, plot the corresponding binding energy as a function of volume and fit the resulting figure to third, fourth and fifth order polynomials to obtain an analytic expression for the dependence. We chose the adequate fit by requiring that the slope of the curve at 100% volume is as close to zero as possible; this occurs for the fourth order fit where the slope is -0.07 GPa (1eV/Å³ = 160.2 GPa, [7]). Once the adequate polynomial is chosen, we can obtain the pressure as follows P = - [∂(BE)/∂(V)] [8] at any point of the adjusted curve. All this extrapolated to 0 K.

The calculations were performed using DMol3 in the Materials Studio suite [9]. To evaluate the BE we used the following parameters: dn basis and an unrestricted spin polarization, a dspp pseudopotential with the VWN functional and a real cut-off radius of 6.0 Å. This cut-off radius was chosen so that the BE of the Wyckoff structure is a *minimum minimorum*. To obtain the eDoS we calculate the electron energy levels both occupied and unoccupied with a smearing factor of 0.2 eV to allow for the smoothing of the density of states. The vDoS is obtained using the finite displacement approach and the results are also smoothed to simulate a bulk sample. The resulting vDoS are normalized to 1 to make the comparison simpler.

In table I we present the coefficients of the fourth order fit and the corresponding $\chi^2$. Table II shows the binding energy as a function of volume and of pressure where the pressure has been determined as the tangent to the fourth order curve adjusted to the simulational points shown in Figure 2.

**Table I.** Three polynomials were adjusted to our simulational results. The fourth order *O(4)* was the most adequate as discussed in the text: $BE(V) = A_0 + A_1 V + A_2 V^2 + A_3 V^3 + A_4 V^4$, where BE is in eV and V in Å³.

|  | O(4) |
|---|---|
| $A_0$ | 2879.50778 |
| $A_1$ | -4.99092 |
| $A_2$ | 3.07680 E-03 |
| $A_3$ | -8.56660 E-07 |
| $A_4$ | 9.1181 E-11 |
| $\chi^2$ | 2.59 E-03 |

**Table II.** Binding energy as function of pressure and percentage of volume. The pressure was determined as the tangent to the curve depicted in Figure 2.

| Volume (Å³) | Volume (%) | Binding Energy (eV) | Pressure (GPa) |
|---|---|---|---|
| 2423.38 | 107 | -193.23047 | -3.11 |
| 2378.08 | 105 | -193.93686 | -2.23 |
| 2332.79 | 103 | -194.44684 | -1.38 |
| 2264.84 | 100 | -194.79269 | -0.07 |
| 2239.02 | 99 | -194.77677 | 0.45 |
| 2196.89 | 97 | -194.55493 | 1.37 |
| 2174.23 | 96 | -194.32712 | 1.90 |
| 2151.60 | 95 | -194.01702 | 2.46 |
| 2128.95 | 94 | -193.61881 | 3.06 |
| 2061.00 | 91 | -191.84727 | 5.11 |
| 2038.35 | 90 | -191.04612 | 5.89 |
| 1993.06 | 88 | -189.08644 | 7.62 |
| 1970.42 | 87 | -188.12609 | 8.58 |
| 1947.76 | 86 | -186.81098 | 9.61 |
| 1925.11 | 85 | -185.35005 | 10.71 |
| 1902.46 | 84 | -183.73161 | 11.89 |
| 1879.81 | 83 | -181.94591 | 13.15 |
| 1857.17 | 82 | -179.98491 | 14.49 |

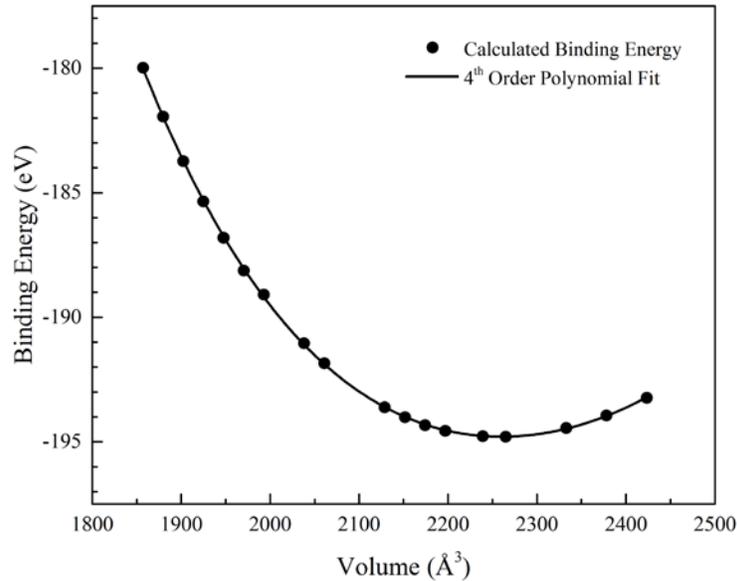

**Figure 2.** Binding energy (BE) obtained computationally for the 18 samples studied, as a function of the volume. The continuous curve is the fourth order fit to the computational points.

Once the compressed (and expanded) supercells were obtained we proceeded to calculate both their vDoS and their eDoS. The results are plotted in Figures 3 and 4, respectively. Two interesting features appear in the vDoS curves; the first one appears at about 2.46 GPa (For pressures of 2.5 GPa, $T_c = 3.9$ K for the Bi-II monoclinic crystalline phase, and for 2.7 GPa, $T_c = 7.2$ K for Bi-III tetragonal phase). The other feature appears at 7.62 GPa (For pressures of 7.7 GPa, $T_c = 8.5$ K for the Bi-V body centered cubic phase). For a pressure of 2.46 GPa a small plateau can be observed at about 8.5 meV and for a pressure of 7.62 GPa a change in curvature can be observed at about 4.6 meV.

For the eDoS the number of states at the Fermi level decreases noticeably in going from a pressure of 7.62 GPa to a value of 8.58 GPa, and this is plotted in Figure 5 where a clearer picture emerges. To convince us of this decrement we did a second run for all samples and the results obtained are essentially the same. Since this pressure is close to 7.7 associated to the corresponding experimental phase change, we believe that the change in eDoS may be associated to the need of the Wyckoff-like structure to relax to a different, more stable phase. Since no other relevant change is observed in this range of pressures, we surmise that the other phases, at 2.5 GPa ($T_c = 3.9$ K) and at 2.7 GPa ($T_c = 7.2$ K) are not as relevant as far as the gain of energy is concerned and therefore do not manifest themselves as prominently.

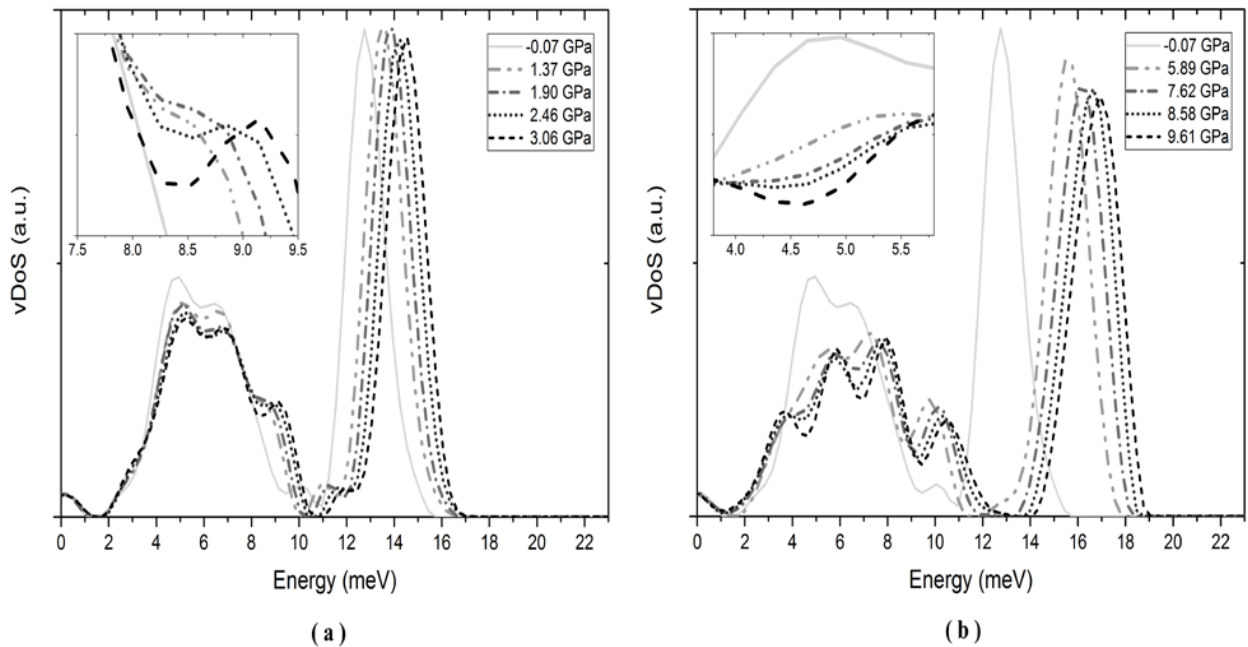

**Figure 3.** Vibrational density of states for some supercells studied in this work. The continuous line corresponds to the structure of Wyckoff. (a) The four curves are in the neighborhood of two possible phase transition, 2.5 and 2.7 GPa. (b) The four curves are in the neighborhood of 7.7 GPa, where another possible phase transition may occur.

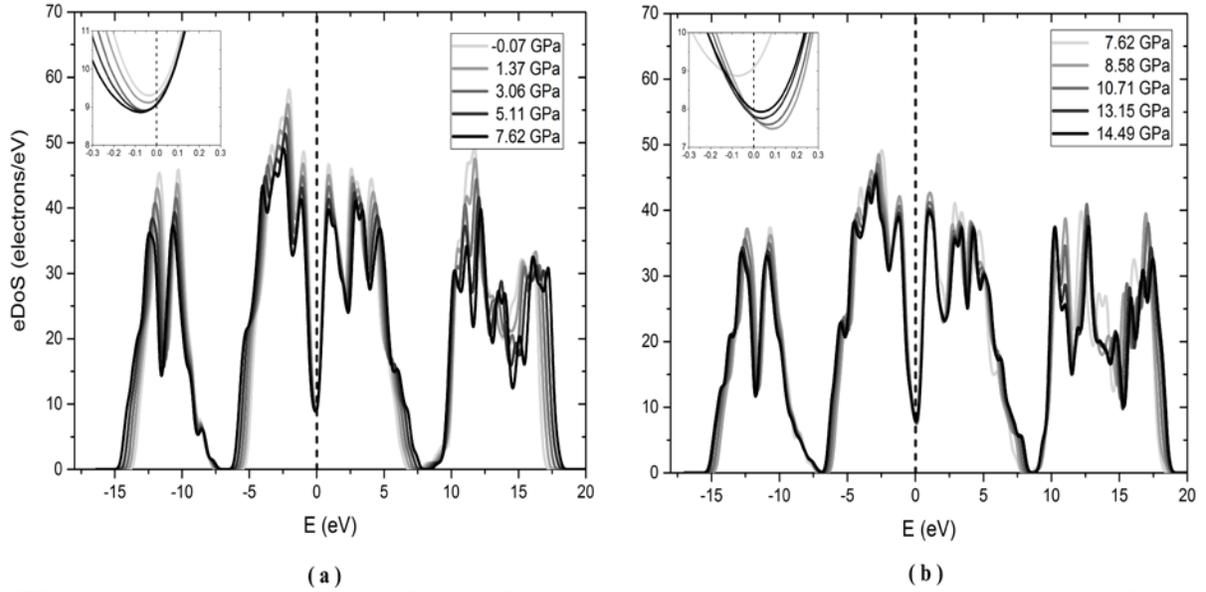

**Figure 4.** Electronic density of states for some supercells studied in this work. (a) The five curves shown have similar N(E$_F$). (b) This graph shows the sudden change in N(E$_F$) in going from 7.62 GPa to 8.58 GPa, in the vicinity of a possible phase transition, 7.7 GPa.

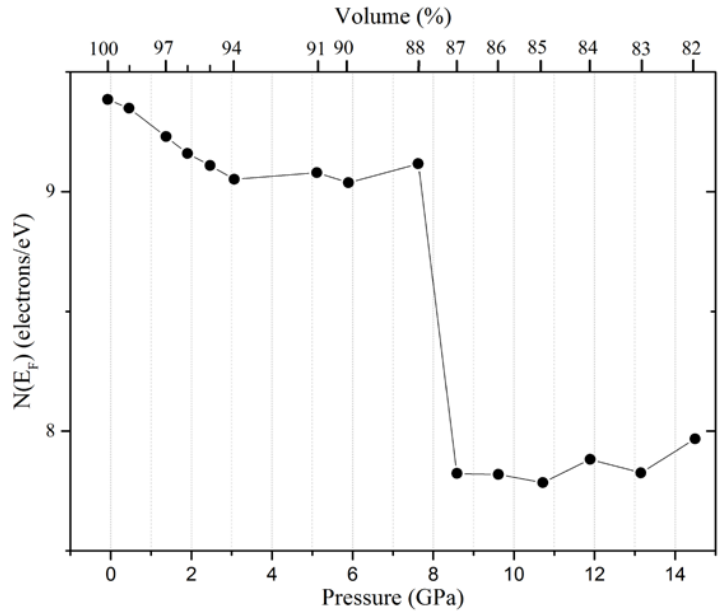

**Figure 5.** Electronic density of states at the Fermi level for 15 of the 18 samples studied. A sudden plunge in the density of states occurs at about 7.7 GPa, where a potential phase transition may occur.

## DISCUSSION

The results found seem to indicate that for pressures in the vicinity of possible phase transitions, changes in the eDoS and the vDoS occur that may indicate the tendency of the Wyckoff-like structures to relax to different more stable structures. However, the evidence is not definite since there is no unequivocal way to correlate these changes to structures that may be unstable. The way to go is to construct other phases and study their stability to discern what structure is to be expected.

Compression induces changes on the extension along the energy axis for the vDoS to the point that the most compressed structure (14.5 GPa) reaches a maximum of about 21 meV while the uncompressed structure has a maximum of 16 meV. This generates an increase in the Debye temperature $\theta_D$ that would have a linear effect in the BCS equation for the superconducting transition temperature $T_c$; however, the drop of the density of states at the Fermi level $N(E_F)$, induces an exponential decrease that overcomes the increase of $\theta_D$. In any case, in order to see how these changes reflect in the superconducting $T_c$, the Wyckoff structure should manifest superconductivity, which it does not, for the low temperatures investigated. Therefore, compressing the crystalline lattice would lower the transition temperature of the structure proposed by Wyckoff and since there is no superconductivity associated to this structure then one can conclude that only increasing the pressure does not make Bi a superconductor. In a previous work [3] we estimated a $T_c$ of 1.3 mK for the stable crystalline lattice; if so, applying pressure to Crystalline Bi would diminish this value since $N(E_F)$ also diminishes.

## CONCLUSIONS

Applying pressure to crystalline bismuth, while maintaining the symmetry properties, generates a diminution of the density of electronic states at the Fermi level that according to BCS would indicate a diminution of the superconducting transition temperature. If in fact the $T_c$ of the crystalline structure is below 1.3 mK compressing this material would make the value of $T_c$ even smaller than 1.3 mK. Clearly, new phases have to be explored and compared with the amorphous properties to understand whether new superconductors can be obtained.

## ACKNOWLEDGMENTS

D.H.R, I.R. and Z.M.P. acknowledge CONACyT for supporting their graduate studies. A.V. R.M.V. and A.A.V. thank DGAPA-UNAM for continued financial support to carry out research projects IN101798, IN100500, IN119105, IN119908, IN112211 and IN110914. M.T. Vázquez and O. Jiménez provided the information requested. Simulations were partially carried out in the Computing Center of DGTIC-UNAM.